\documentclass[paper a4]{article}
\title{Quantum mechanics on noncommutative plane and sphere
from constrained systems.}
\author{A.A. Deriglazov\footnote{alexei@fisica.ufjf.br ~ On leave of
absence from Dept. Math. Phys., Tomsk Polytechnical University,
Tomsk, Russia.}}
\date{Dept. de F\'\i sica, ICE, Universidade Federal de Juiz de Fora,\\
MG, Brasil.}
\begin{document}
\maketitle
\large
\begin{abstract}
It is shown that quantum mechanics on noncommutative (NC) spaces
can be obtained by canonical quantization of some underlying
constrained systems.
Noncommutative geometry arises after taking into account the second
class constraints presented in the models.
It leads, in particular, to a possibility of
quantization in terms of the initial NC variables.
For a two-dimensional plane we present two Lagrangian actions,
one of which admits addition of an arbitrary potential.
Quantization leads to quantum mechanics with ordinary product replaced
by the Moyal product. For a three-dimensional case we present
Lagrangian formulations for a particle on NC sphere as well as for
a particle on commutative sphere with a magnetic monopole at the center,
the latter is shown to be equivalent to the model of usual rotor.
There are several natural possibilities to choose physical
variables, which lead either to commutative or to NC brackets for
space variables.
In the NC representation all information on the
space variable dynamics is encoded in the NC geometry. Potential
of special form can be added,
which leads to an example of quantum mechanics on the NC sphere.
\end{abstract}

{\bf PAC codes:} 0460D, 1130C, 1125 

\noindent
\section{Introduction and symmary.}
Recently quantum mechanics on noncommutative spaces (NQM) have received a
considerable discussion [1-7]. For two-dimensional plane it can be
characterized by the following brackets
($\epsilon_{ab}=-\epsilon_{ba},$ ~  $a, b=1,2, ~ \epsilon_{12}=1$)
\begin{eqnarray}\label{1}
\{x_a, x_b\}=\theta\epsilon_{ab}, \qquad
\{x_a, p_b\}=\delta_{ab}, \qquad
\{p_a, p_b\}=0,
\end{eqnarray}
and by the Hamiltonian $H=\frac{p^2}{2m}+V(x)$
with some potential $V(x)$. To make this situation tractable, the
prescription is to consider new variables
\begin{eqnarray}\label{3}
\tilde x_a=x_a+\frac{\theta}{2}\epsilon_{ab}p_b, \qquad \tilde p_a=p_a,
\end{eqnarray}
which obey the canonical brackets and thus can be quantized in the
standard way. It leads to the
Schr\"{o}dinger
equation
\begin{eqnarray}\label{4}
E\Psi(\tilde x)=\frac{1}{2m}\tilde p^2\Psi(\tilde x)+
V(\tilde x_a-\frac{\theta}{2}\epsilon_{ab}\tilde p_b)\Psi(\tilde x),
\end{eqnarray}
where the last term can be rewritten [8, 9, 3] through the Moyal product
\begin{eqnarray}\label{5}
V(\tilde x_a-\frac{\theta}{2}\epsilon_{ab}\tilde p_b)\Psi(\tilde x)=
V(\tilde x)\star\Psi(\tilde x).
\end{eqnarray}
Thus one obtains quantum mechanics in terms of the commutative variables
$\tilde x, \tilde p$, but  with the ordinary product replaced by the star
product.

Let us recall that in some cases appearance of the noncommutative
geometry [10] has a natural interpretation as
resulting from the canonical
quantization of some underlying constrained system. In particular, this
interpretation is possible for the lowest level Landau problem [11, 9]
and for the open string in a B-field background [12-14]. Regarding the
NC quantum mechanics on the plane, one possibility was proposed in [1],
starting from higher
derivative mechanical action. It leads to the NC particle
with extra physical degrees of freedom. In this relation it is natural
to ask whether a similar interpretation is possible for NC quantum
mechanics of the scalar particle (\ref{1})-(\ref{5}), as well as for
the corresponding generalization on three dimensional sphere. Here we
demonstrate that it is actually the case.
Our starting point will be some mechanical system (without higher
derivatives) formulated in
an appropriately extended configuration space. Nonphysical character of
the corresponding extra degrees of freedom is supplied by second
class constraints.
The noncommutative geometry (\ref{1})
arises after introduction the Dirac bracket corresponding to these
constraints, while the
prescription (\ref{3}) becomes, in fact, the standard necessary step
for the canonical quantization of a system with second class constraints
[15, 16].

The work is organised as follows. In Sect. 2 we discuss models which
lead to quantum mechanics
on the NC plane. The Lagrangian action, which is
appropriate for at most quadratic potential, looks as follows
\begin{eqnarray}\label{6}
S=\int d\tau\left[-\frac{m}{2}\dot v^2+2(\dot v-\dot x)\theta^{-1}v
-\frac{2}{m\bar\theta^2}v^2-U(x)\right],
\end{eqnarray}
where $x_a(\tau), ~ v_a(\tau)$ are the configuration space variables and
$\theta_{ab}=\bar\theta\epsilon_{ab}$. The variables $v_a$ are subject
to the second class constraints  and can be omitted from
the consideration after the Dirac bracket is
introduced. The physical sector consist of $x_a$ and the conjugated
momentum $p_a$. The Dirac bracket for $x_a$ turns out to be nontrivial,
with the noncommutativity parameter being
$\theta=\bar\theta{\det}^{-1}(1-\frac{m\bar\theta^2}{4}\partial\partial U)$.
The parameter (and rank of matrix of the constraint
brackets) depend on the potential.
It explains appearance of two phases [4-6] of the resulting NQM:
for a critical
value of the parameter, the model involves first class constraints instead
of the second class ones.
For the case $\partial_a\partial_b U=const$ one can easily find the
canonical variables, see Eq.(\ref{22}).
Quantization leads to NQM (\ref{4}) with the potential
$V=-\frac{m\bar\theta^2}{8}\partial_aU\partial_aU+U$. For an arbitrary
potential $U(x)$, the noncommutativity parameter $\theta$ depends on
$x_a$ and one is faced with the problem of diagonalization of the
brackets, Eq.(\ref{21}) below. Surprisingly enough, the
problem can be resolved
if one starts from the action, which is obtained from (\ref{6})
omitting the first term
\begin{eqnarray}\label{9}
S=\int d\tau\left[2(\dot v-\dot x)\theta^{-1}v-\frac{2}{m\bar\theta^2}v^2-
V(x)\right].
\end{eqnarray}
It can be considered as the action of ordinary particle (with position
$x_a$) written in the first order form, with the
``Chern-Simons term'' for $v$ added: $\dot v\theta^{-1}v$.
The action is similar to the one discussed by Lukierski at all [1], but
does not involve of higher derivatives. As a concequence, there are no
``internal'' oscillator modes in the physical sector. Below
we show that this action leads to NQM of the scalar particle
(\ref{1})-(\ref{5}) with the same potential $V$.

In Sect. 3 we demonstrate that the same procedure works for
noncommutative sphere in three dimensions [19-21]. We propose the
following action
\begin{eqnarray}\label{201}
S=\int d\tau\left[-\epsilon_{ijk}\dot v_iv_jx_k-
v^2+\phi(x_i^2-1)-V(v^2)\right],
\end{eqnarray}
where $\epsilon_{ijk}=\epsilon_{[ijk]}, ~ \epsilon_{123}=1$. The variables
$v_i$ play the same role as in the previous case (\ref{6}). The variables
$x_i$ are restricted to lie on the sphere $x^2=1$. The kinematic constraint
is included into the action by using of the Lagrangian multiplier $\phi$.
Dynamics is governed by second order differential equations,
which is supplied by the presence of the term $v^2$.
We restrict ourselves to
$SO(3)$-invariant potential $V(v^2)$. The combination $x_iv_i$ is not
included into the potential since it would lead to deformation of the
constraint system algebra as compared with the free case $V=0$
(see below).
In the Hamiltonian formalism
the essential constraints of the model are
\begin{eqnarray}\label{102}
G_i\equiv p_i+\epsilon_{ijk}v_jx_k=0, \quad
\pi_i=0, \quad
x_i^2-1=0, \quad x_iv_i=0,
\end{eqnarray}
where $(x, ~ \pi)$ and $(v, ~ p)$ form canonical pairs (in these notations
commutative relations appear in the standard form, see Eq.(\ref{222})).
The constraints form the second class system.
The corresponding Dirac bracket is constructed
and brackets for the phase space variables are presented in $SO(3)$
covariant form. Using the constraints $G_i=0$ one can represent one of the
variables $(x, ~ v, ~ p)$ through the remaining ones,
which leads either to commutative or to NC brackets for space variables.
The representations are discussed in Sect. 4.
$(x, ~ p)$-representation is characterized by NC space geometry and
trivial dynamics for the corresponding space variables. In
$(v, ~ p)$-representation the geometry can be made commutative by
transition to the canonical variables,
the dynamics of which is governed then by
nonlinear equations. In Sect. 5 we present and discuss slight
modification of the action (\ref{201}) which describes a particle on the
commutative sphere with a monopole at the center. In particular, we
show canonical equivalence of this model and model of the rotor.
In the end of the section some possible generalizations of the
action (\ref{201}) are discussed.

\section{Particle on the noncommutative plane.}

Starting from the action (\ref{6}), one finds in the Hamiltonian
formalism the primary constraints
\begin{eqnarray}\label{11}
G_a\equiv p_a+2\theta^{-1}_{ab}v_b=0,
\end{eqnarray}
and the Hamiltonian
\begin{eqnarray}\label{16}
H=-\frac{1}{2m}\pi^2+\frac{2}{m}\pi\theta^{-1}v+U(x)+
\lambda(p+\theta^{-1}y).
\end{eqnarray}
Here $p, ~ \pi$ are conjugated momenta for $x, ~ v$ and $\lambda$ is the
Lagrangian multiplier for the constraint. Further analysis gives
the secondary constraints
\begin{eqnarray}\label{17}
\dot G_a=0 \Longrightarrow
T_{a}\equiv\pi_a-2\theta^{-1}_{ab}v_b+\frac{m}{2}\theta_{ab}\partial_bU=0,
\end{eqnarray}
as well as equations for determining the Lagrangian multipliers
\begin{eqnarray}\label{18}
F\lambda=-\frac{2}{m}(\pi-\theta^{-1}v), \qquad
F_{ab}\equiv\delta_{ab}-\frac{m\bar\theta^2}{4}\partial_a\partial_b U.
\end{eqnarray}
Next step depends on the rank of the matrix $F$. If $\det F=0$, the model
involves first class constraints (see also Eq.(\ref{20})), which explains
appearance of two phases [4-6] of the resulting NQM. Let us consider the
nondegenerated case $\det F\ne 0$.
Then the constraints form the second class system
\begin{eqnarray}\label{20}
\{G_a, G_b\}=0, \qquad \{T_a, T_b\}=-4\theta^{-1}_{ab}, \qquad
\{G_a, T_b\}=2F_{ac}\theta^{-1}_{cb}.
\end{eqnarray}
Introducing the Dirac bracket
\begin{eqnarray}\label{12}
\{A, B\}_D=\{A, B\}-\{A, G\}\theta\triangle^{-1}\{G, B\}- \cr
\{A, G\}\frac{1}{2}F^{-1}\theta\{T, B\}-
\{A, T\}\frac{1}{2}\theta F^{-1}\{G, B\},
\end{eqnarray}
the variables $v, \pi$ can be omitted from consideration, while
for the remaining physical variables $x, ~ p$ one obtains from
Eq.(\ref{12}) the following brackets:
\begin{eqnarray}\label{21}
\{x_a, x_b\}=\triangle^{-1}\theta_{ab}, \quad
\{x_a, p_b\}=F^{-1}_{ab},
\quad \{p_a, p_b\}=0.
\end{eqnarray}
The noncommutativity parameter depends on the potential through the
quantity
\begin{eqnarray}\label{22}
\triangle\equiv\det(1-\frac{m\bar\theta^2}{4}\partial\partial U).
\end{eqnarray}
Let us restrict ourselves to the case $\partial_a\partial_b U=const$.
To quantize the system
one needs to find the canonical variables [16], which in this case
turn out to be
\begin{eqnarray}\label{22}
\tilde x_a=F_{ab}x_b+\frac{1}{2}\theta_{ab}p_b, \quad
\tilde p_a=p_a.
\end{eqnarray}
They obey the standard brackets
$\{\tilde x_a, \tilde x_b\}=0, ~ \{\tilde x_a, \tilde p_b\}=\delta_{ab},
~ \{\tilde p_a, \tilde p_b\}=0$.
The Hamiltonian in terms of the canonical variables is
(the term $F$ can be equally included into the kinetic part
of the Hamiltonian [26])
\begin{eqnarray}\label{23}
H_{ph}=\frac{1}{2m}\tilde p^2-
\frac{m\bar\theta^2}{8}\partial_aU\partial_aU|_{x(\tilde x, \tilde p)}+
U[F^{-1}(\tilde x-\frac{1}{2}\theta\tilde p)],
\end{eqnarray}
where the term with derivatives of the potential comes from Eq.(\ref{17}).
The resulting system can be quantized now in the standard way. Note that
the underlying potential $U$ and the final one turn out to be different
for this model. For example, starting from the harmonic oscillator
$U=\frac{k}{2}|x|^2$, one obtains the NQM which corresponds to
oscillator with
renormalized rigidity $\tilde k=(1-\frac{m\bar\theta^2k}{4})^{-1}k$,
namely
\begin{eqnarray}\label{24}
V=
[-\frac{m\bar\theta^2}{8}\partial_aU\partial_aU+
U]|_{x(\tilde x, \tilde p)}=\frac{\tilde k}{2}
|\tilde x-\frac{1}{2}\theta\tilde p|^{2}.
\end{eqnarray}
Note also that in absence of the potential ($U=0$) the model (\ref{6})
describes the free NC particle which is characterised
by the equations of motion $\dot x_a=\frac{1}{m}p_a, ~ \dot p_a=0$
and by the relations (\ref{1}).

Let us return to the case of an arbitrary potential. As it was mentioned,
the complicated brackets (\ref{21}) arise due to the fact that the
secondary constraints (\ref{17}) involve derivative of the potential.
While existence of the canonical variables is guaranteed
by the known theorems [16], it is problematic to find a solution in the
manifest form.
One possibility to avoid the problem is to construct action which will
create the primary constraints only. Since $U(x)$ does not contain
the time derivative, it can not give contribution into the primary
constraints. An appropriate action is\footnote{An equivalent form of
the action can be obtained by the shift: $x \to x'=x-v$.}
\begin{eqnarray}\label{25}
S=\int d\tau\left[2(\dot v-\dot x)\theta^{-1}v-\frac{2}{m\bar\theta^2}v^2-
V(x)\right],
\end{eqnarray}
where $x_a, ~ v_a$ are the configuration space variables. Configuration
space dynamics is governed by the
second order equations which is supplied
by the term $v^2$.
Following the Dirac procedure one obtains the primary second
class constraints
\begin{eqnarray}\label{26}
G_a\equiv p_a+2\theta^{-1}_{ab}v_b=0, \qquad
T_{a}=\pi_a-2\theta^{-1}_{ab}v_b=0,
\end{eqnarray}
and the Hamiltonian
\begin{eqnarray}\label{27}
H=\frac{2}{m\bar\theta^2}v^2+V(x)+\lambda_1G+\lambda_2T.
\end{eqnarray}
Remaining analysis is similar to the previous case.
Introducing the Dirac bracket (\ref{12}) (where $F=1$ now) , the
variables $v, ~ \pi$ can
be omitted, while for $x, ~ p$ one has the brackets (\ref{1}). Defining
the canonical variables
$\tilde x_a=x_a+\frac{1}{2}\theta_{ab}p_b, ~ \tilde p_a=p_a$,
one obtains the physical Hamiltonian
$H=\frac{2}{m}\tilde p^2+V(\tilde x-\frac{1}{2}\theta\tilde p)$,
thus reproducing the NQM (\ref{4}), (\ref{5}) for the case of an
arbitrary potential.

We have demonstrated that quantum mechanics on NC plane
can be considered as resulting from direct canonical
quantization of the underlying constrained systems (\ref{6}), (\ref{9}).
It implies, that
instead of the star product (\ref{4}), (\ref{5}), one can equally use now
other possibilities to quantize the system. In particular, the
conversion scheme [17] or the embedding formalism [18] can be applied.
For example, it is not difficult to rewrite the formulation (\ref{25})-
(\ref{27}) as a first class constrained system. Namely, let us keep
$G$-constraint only and define the deformed Hamiltonian as
\begin{eqnarray}\label{30}
\tilde H=\frac{2}{m\bar\theta^2}v^2+V[x-\frac{1}{2}\theta
(\pi-2\theta^{-1}v)]+\lambda G.
\end{eqnarray}
Since $\{G, \tilde H\}=0$, it is equivalent formulation of the problem
(\ref{27}), the latter is reproduced in the gauge $T=0$. Now one can
quantize all the variables canonically, while the first class constraint
$G=0$ can be imposed as restriction on the wave function. It implies
quantization in terms of the initial NC variables.
Another possibility is to consider the gauges different from $T=0$.
For example, one can take $\pi=0$, which can lead to simplification
of the eigenvalue problem (\ref{4}).

\section{Particle on the noncommutative sphere.}

Here we show that dynamics on the NC sphere can be described
in a similar fashion, starting from the action (\ref{201}).
From manifest form of the action it follows that velocities
do not enter
into expressions for definition of conjugated momentum in the
Hamiltonian formulation. On the first stage of the Dirac procedure
one finds the primary constraints
\begin{eqnarray}\label{202}
G_i\equiv p_i+\epsilon_{ijk}v_jx_k=0, \quad
T_i\equiv\pi_i=0, \quad p_{\phi}=0,
\end{eqnarray}
where $p_i$ are conjugated momentum for $v_i$
while $\pi_i$ corresponds to
$x_i$. The Hamiltonian is
\begin{eqnarray}\label{203}
H=v^2-\phi(x_i^2-1)+V(v^2)+\lambda_iG_i+
\bar\lambda_iT_i+\lambda p_{\phi},
\end{eqnarray}
where $\lambda$ are the Lagrangian multipliers for the corresponding
constraints. The constraints obey the following Poisson bracket algebra
\begin{eqnarray}\label{204}
\{G_i, G_j\}=2\epsilon_{ijk}x_k, \quad
\{G_i, T_j\}=-\epsilon_{ijk}v_k, \quad
\{T_i, T_j\}=0.
\end{eqnarray}
Matrix composed from the brackets admits two null-vectors
$w_1=(0,v_j), ~ w_2=(v_i, -2x_j)$, so the system $(G, ~ T)$ involve two
first class constraints at this stage: $v_iT_i$ and $v_iG_i-2x_jT_j$.
From (\ref{202}) one has the consequences
$v_ip_i=0, ~  x_ip_i=0$.
At the second stage of the Dirac procedure
there appear the equations
\begin{eqnarray}\label{206}
\dot p_{\phi}=0 ~ \Longrightarrow ~ x_i^2-1=0, \cr
\dot G_i=0 ~ \Longrightarrow ~ -2(1+V')v_i+
2\epsilon_{ijk}\lambda_jx_k-\epsilon_{ijk}\bar\lambda_jv_k=0, \cr
\dot T_i=0 ~ \Longrightarrow ~
2\phi x_i-\epsilon_{ijk}\lambda_jv_k=0,
\end{eqnarray}
where $V'\equiv\frac{\partial V}{\partial v^2}$. From these equations
one extracts three secondary constraints
\begin{eqnarray}\label{207}
S\equiv x_i^2-1=0, \quad \bar S\equiv x_iv_i=0, \quad
\Phi\equiv\phi+\frac 12v_i^2(1+V')=0,
\end{eqnarray}
while the remaining equations involve the Lagrangian multipliers. They
will be resolved in the manifestly $SO(3)$-covariant form below.
On the next step there arise equations for the Lagrangian
multipliers only
\begin{eqnarray}\label{208}
\dot\Phi=0 ~ \Longrightarrow ~ \lambda+\{\Phi, G_j\}\lambda_j+
\{\Phi, T_j\}\bar\lambda_j=0, \cr
\dot S=0 ~ \Longrightarrow ~ x_i\bar\lambda_i=0, \qquad
\dot{\bar S}=0 ~ \Longrightarrow ~ v_i\bar\lambda_i+x_j\lambda_j=0,
\end{eqnarray}
which finishes the Dirac procedure for revealing the constraints.
To determine the Lagrangian multipliers one has now equations
(\ref{206}), (\ref{208}). Their consequences are
$x_i\lambda_i=
x_i\bar\lambda_i=
v_i\lambda_i=
v_i\bar\lambda_i=0$.
Using these equations, one resolves Eqs.(\ref{206}), (\ref{208}) as
\begin{eqnarray}\label{210}
\lambda_i=(1+V')p_i, \qquad \bar\lambda_i=\lambda=0.
\end{eqnarray}
The Hamiltonian equations of motion for the model can be obtained with
the help of Eqs.(\ref{203}), (\ref{210}).
They will be discussed in the next section.
Since all the multipliers have been determined, the constraints
(\ref{202}), (\ref{207}) form the second class system and thus can be
taken into account by transition to the Dirac bracket.
After introduction of the Dirac bracket corresponding to the pair
$p_{\phi}=0, ~ \Phi=0$, the variables $\phi, ~ p_{\phi}$ can be
omitted from
consideration. The Dirac brackets for the remaining variables coincide
with the Poison one. To find the Dirac bracket which corresponds
to the remaining eight constraints one
needs to invert $8\times 8$ matrix composed from
Poisson brackets of these constraints. To simplify the problem,
we prefer to do this in two steps: first, let us construct an
intermediate Dirac
bracket which corresponds to the constraints $G_a=0, ~ T_a=0, ~
a=1,2$, and then bracket which corresponds to the remaining
constraints $G_3, ~ T_3, ~ S, ~ \bar S$. Consistency of this procedure is
guaranteed by the known theorems [16]. On the first step one has the
Poisson brackets
\begin{eqnarray}\label{211}
\{G_a, G_b\}=2\epsilon_{ab}x_3, \quad
\{G_a, T_b\}=-\epsilon_{ab}v_3, \quad
\{T_a, T_b\}=0.
\end{eqnarray}
Then the intermediate Dirac bracket is
\begin{eqnarray}\label{212}
\{A, B\}_{D1}=\{A, B\}-\{A, G_a\}\frac{\epsilon_{ab}}{v_3}\{T_b, B\}- \cr
\{A, T_a\}\frac{\epsilon_{ab}}{v_3}\{G_b, B\}-
\{A, T_a\}\frac{2x_3}{v_3^2}\epsilon_{ab}\{T_b, B\}.
\end{eqnarray}
Now one can use the equations $G_a=0, ~ T_a=0$ in any expression.
As a consequence, the remaining constraints can be taken in the form
\begin{eqnarray}\label{213}
G_3\equiv x_ip_i=0, \qquad
T_3\equiv\pi_3=0, \cr
S\equiv x_i^2-1=0, \qquad
\bar S\equiv\frac{1}{x_3}(v_3+J_3)=0,
\end{eqnarray}
and obey the  $D1$-algebra
\begin{eqnarray}\label{214}
\{G_3, S\}_{D1}=-\frac{4x_3}{J_3}, \quad \{G_3, \bar S\}_{D1}
=-2\left(1+\frac{p_a^2}{J_3^2}\right), \cr
\{T_3, S\}_{D1}=0, \quad
\{T_3, \bar S\}_{D1}=-\frac{J_3^2-p_a^2}{x_3^2J_3}, \quad
\{S, \bar S\}_{D1}=\frac{4x_3p_3}{J_3^2},
\end{eqnarray}
where $J_i$ are the rotation generators: $J_i\equiv\epsilon_{ijk}x_jp_k$.
The corresponding matrix can be easily inverted, and the final expression
for the Dirac bracket is
\begin{eqnarray}\label{215}
\{A, B\}_{D}=\{A, B\}-
\{A, G_3\}\frac{x_3^2p_3}{J_3^2-p_a^2}\{T_3, B\}- \cr
\{A, G_3\}\frac{J_3}{4x_3}\{S, B\}+
\{A, T_3\}\frac{x_3^2p_3}{J_3^2-p_a^2}\{G_3, B\}+\cr
\{A, T_3\}\frac{x_3(J_3^2+p_a^2)}{2(J_3^2-p_a^2)}\{S, B\}-
\{A, T_3\}\frac{x_3^2J_3}{J_3^2-p_a^2}\{\bar S, B\}+ \cr
\{A, S\}\frac{J_3}{4x_3}\{G_3, B\}-
\{A, S\}\frac{x_3(J_3^2+p_a^2)}{2(J_3^2-p_a^2)}\{T_3, B\}+ \cr
\{A, \bar S\}\frac{x_3^2J_3}{J_3^2-p_a^2}\{T_3, B\},
\end{eqnarray}
where all the brackets on the r.h.s. are $D1$-brackets.
Note that the complete constraint system (\ref{202}), (\ref{207})
is $SO(3)$-covariant. Consequently, one
expects that the final expressions for the brackets can be rewritten in
$SO(3)$-covariant form also. It is actually the case. For example,
from Eq.(\ref{215}) one obtains for the variables $x_i$
\begin{eqnarray}\label{216}
\{x_1, x_2\}_{D}=-\frac{x_3}{J_3^2}\left[2+
\frac{x_3^2(2p_i^2+J_3^2)}{J_3^2-p_a^2}\right], \cr
\{x_a, x_3\}_{D}=-\frac{1}{J_3p_i^2}
\left[\frac{2x_3p_3}{J_3}\epsilon_{ac}p_c+2p_a+(J_3^2+2p_a^2)
\epsilon_{ac}x_c\right].
\end{eqnarray}
Using the equalities
\begin{eqnarray}\label{217}
J^2\equiv J_i^2=p_i^2=v_i^2, \qquad J_3^2-p_a^2=-x_3^2p_i^2, \cr
x_3p_3p_a-J_3\epsilon_{ab}p_b+p_b^2x_a=0,
\end{eqnarray}
which are true on the constraint surface (\ref{202}), (\ref{207}), the
equations (\ref{216}) can be presented in $SO(3)$-covariant form
$\{x_i, x_j\}_{D}=\frac{1}{J^2}\epsilon_{ijk}x_k$, ~  $i, j=1,2,3$.
Other brackets can be computed from Eq.(\ref{215}) in a similar fashion.
After tedious calculations one obtains the following result
\begin{eqnarray}\label{219}
\{x_i, x_j\}=\frac{1}{J^2}\epsilon_{ijk}x_k, \qquad
\{x_i, p_j\}=\frac{1}{J^2}J_ix_j, \cr
\{p_i, p_j\}=-\frac{1}{2}\epsilon_{ijk}x_k,
\end{eqnarray}
\begin{eqnarray}\label{220}
\{v_i, v_j\}=-\frac{1}{2}\epsilon_{ijk}x_k, \qquad
\{v_i, x_j\}=-\frac{1}{J^2}x_ip_j, \cr
\{v_i, p_j\}=\frac{1}{2}(\delta_{ij}+x_ix_j).
\end{eqnarray}
Since $\{x_i, J^2\}=0$, the operator $J^2$ can be included into
redefinition of $x_i$: $\tilde x_i\equiv J^2x_i$, then $\tilde x_i$
obeys $SU(2)$ algebra
$\{\tilde x_i, \tilde x_j\}=\epsilon_{ijk}\tilde x_k$, and
is constrained to lie on the
fuzzy sphere $\tilde x^2=(J^2)^2$. Quantum realization and
irreducible representations of
such a kind of an algebraic structure were
considered, in particular, in [20, 21]. One notes that the algebra
obtained (\ref{219}) has much simpler structure as
compared with the one proposed in [20] from algebraic considerations.

Since the second class constraints were taken into account, one can
now use them in any expression. In particular, from Eqs.(\ref{202}),
(\ref{207}) it follows that as the physical sector variables one can
choose either $(x_i, p_i)$ or $(v_i, p_i)$, or $(x_i, v_i)$.
Relation between these representations is given by the first equation
from (\ref{202}), which can be written in one of the following
forms\footnote{The representations for NC plane can be obtained
in a similar fashion starting from Eq.(\ref{26}), and are not interesting
due to linear character of the constraints.}
\begin{eqnarray}\label{221}
p_i=-\epsilon_{ijk}v_jx_k, \qquad
v_i=-\epsilon_{ijk}x_jp_k, \qquad
x_i=\frac{1}{v^2}\epsilon_{ijk}v_jp_k.
\end{eqnarray}
Let us point out that for any given choice, the remaining nonphysical
variable looks formally as the rotation generator in the corresponding
representation. The equations (\ref{221}) relate different
representations of the particle
dynamics on NC sphere which are discussed in the next section.

\section{Three representations for the particle dynamics on
noncommutative sphere.}
To discuss classical dynamics of the particle on NC sphere it will be
sufficient to consider the free case $V=0$. In what follows, we will
preserve SO(3) covariance which implies that two of the constraints are not
resolved in the manifest form. Note also that the
variables $\phi, ~ p_{\phi}$ are trivially constrained $\phi=0, ~
p_{\phi}=0$ and thus are omitted from consideration.

{\bf Noncommutative $(x_i, ~ p_i)$-representation.}
Taking $x, ~ p$ as the basic variables, their algebra is
\begin{eqnarray}\label{222}
\{x_i, x_j\}=\frac{1}{J^2}\epsilon_{ijk}x_k, \quad
\{x_i, p_j\}=\frac{1}{J^2}J_ix_j, \quad
\{p_i, p_j\}=-\frac{1}{2}\epsilon_{ijk}x_k.
\end{eqnarray}
Equations of motion follow from (\ref{203}), (\ref{210})
\begin{eqnarray}\label{223}
\dot x_i=0, \qquad \dot p_i=\epsilon_{ijk}x_jp_k,
\end{eqnarray}
and are accompanied by two constraints
\begin{eqnarray}\label{224}
x_i^2-1=0, \qquad x_ip_i=0.
\end{eqnarray}
The physical Hamiltonian has the form $H_{ph}=p^2$.
One notes that $\{x_i, J^2\}=0$, so $J^2$ can be absorbed into redefinition
of $x_i: \tilde x_i\equiv J^2x_i$. The algebra acquires then the form
\begin{eqnarray}\label{225}
\{\tilde x_i, \tilde x_j\}=\epsilon_{ijk}\tilde x_k, \quad
\{\tilde x_i, p_j\}=\epsilon_{ijk}p_k, \quad
\{p_i, p_j\}=-\frac{1}{2J^2}\epsilon_{ijk}\tilde x_k.
\end{eqnarray}
{\bf Commutative $(v_i, ~ p_i)$-representation.}
In this case the bracket algebra is
\begin{eqnarray}\label{226}
\{v_i, v_j\}=-\frac{1}{2p^2}v_{[i}p_{j]}, \qquad
\{v_i, p_j\}=\frac{1}{2}(\delta_{ij}-\frac{v_iv_j+p_ip_j}{v^2}), \cr
\{p_i, p_j\}= -\frac{1}{2p^2}v_{[i}p_{j]}.
\end{eqnarray}
Dynamics turns out to be nontrivial for both variables
\begin{eqnarray}\label{227}
\dot v_i=p_i, \quad \dot p_i=-v_i, \quad
v_i^2=p_i^2, \quad v_ip_i=0,
\end{eqnarray}
which implies $\ddot v+v=0$ for the configuration space variable.
Note that the representation turns out to be symmetric under the change
$v\rightarrow -p, ~~ p\rightarrow v$.

Let us compare these two representations of the particle dynamics on
NC sphere. Since the NC geometry has been obtained by using the
Dirac bracket, there exists transformation to new variables,
in terms of which the bracket acquires the canonical form [16].
The corresponding theorem
states that constraints of a theory become a part of the new variables
after this transformation. Being applied to the case under consideration,
it means that the new variables will have the following structure:
\begin{eqnarray}\label{229}
(x_i, ~ \pi_i, ~ v_i, ~ p_i) \Longrightarrow (\tilde x_i=x_i-
\frac{\epsilon_{ijk}v_jp_k}{v^2}, ~ \pi_i, ~ \tilde v_3, ~ \tilde p_3,
~ \tilde v_a, ~ \tilde p_a),
\end{eqnarray}
where $\tilde v_a, ~ \tilde p_a$ are the
physical variables with the canonical brackets, in particular:
$\{\tilde v_a, \tilde v_b\}=0$.  From the expression
(\ref{202}), (\ref{229}) one
notes that the theorem naturally selects $(v,~ p)$-representation for
transition to the canonical brackets, which is the reason for the name:
``commutative representation''. From Eq.(\ref{227}) it follows that
the canonical coordinates have nontrivial equations of motion
(see also the next section). In contrast, in the NC
$(x, ~ p)$-representation the configuration space dynamics (\ref{223})
turns out to be trivial. Thus, the NC description
implies that all information on the dynamics is encoded in
NC geometry. Similar situation was observed for
$SO(n)$ nonlinear sigma-model in [22] and for the Green-Schwarz
superstring in the covariant gauge in [23].

{\bf $(x_i, ~ v_i)$-representation} coincides with
$(x_i, p_i)$-representation.

\section{Particle on commutative sphere with a magnetic monopole at the
center and the Rotor.}
In this section we show that slight modification of the action
(\ref{201}) gives description for a particle with a monopole at the
center of the sphere [24]. It will be demonstrated also that
this model is equivalent to the model of usual rotor.

Let us consider the action (\ref{201}) with the variables $x$ and $v$
interchanged in the first term
\begin{eqnarray}\label{230}
S=\int d\tau\left[-\epsilon_{ijk}\dot x_ix_jv_k-
v^2+\phi(x_i^2-1)-V(v^2)\right].
\end{eqnarray}
Canonical momentum for $x_i$ is denoted through $p_i$ while
$\pi_i$ corresponds to the variable $v_i$ (the notations are opposite
to the ones adopted for the model (\ref{201})).
In these notations analysis of the model turns out to be similar
to the previous case, so we present the final results only. The essential
constraints of the theory are
\begin{eqnarray}\label{231}
G_i\equiv p_i+\epsilon_{ijk}x_jv_k=0, \quad
T_i\equiv\pi_i=0, \cr
S\equiv x_i^2-1=0, \quad \bar S\equiv x_iv_i=0,
\end{eqnarray}
and can be taken into account by transition to the Dirac bracket.
After that, dynamics of the model can be presented in one of the
following three forms.

{\bf $(x_i, v_i)$-representation.} In terms of these variables the
bracket algebra is
\begin{eqnarray}\label{232}
\{x_i, x_j\}=0, \quad
\{x_i, v_j\}=\epsilon_{ijk}x_k, \quad
\{v_i, v_j\}=\epsilon_{ijk}v_k,
\end{eqnarray}
while their dynamics is governed by the equations (free case)
\begin{eqnarray}\label{233}
\dot x_i=-2\epsilon_{ijk}x_jv_k, \quad
\dot v_i=0, \quad
x_i^2-1=0, \quad x_iv_i=0.
\end{eqnarray}
For the physical Hamiltonian one has the expression (remember that $v$
is noncommutative variable)
\begin{eqnarray}\label{234}
H_{ph}=v^2+V(v^2).
\end{eqnarray}
The algebra obtained (\ref{232}) corresponds to the particle on
commutative sphere with a monopole at the center (note the relations
(\ref{217}))[24].

{\bf $(x_i, ~ p_i)$-representation.}
In this case one has the brackets
\begin{eqnarray}\label{235}
\{x_i, x_j\}=0, \quad
\{x_i, p_j\}=\delta_{ij}-x_ix_j, \quad
\{p_i, p_j\}= -(x_ip_j-x_jp_i).
\end{eqnarray}
Equations of motion turn out to be nontrivial for both variables
\begin{eqnarray}\label{236}
\dot x_i=2p_i, \quad \dot p_i=-2p^2x_i, \quad
x_i^2-1=0, \quad x_ip_i=0.
\end{eqnarray}
which implies $\ddot x_i+4p^2x_i=0$ for the configuration space variables.
The equations (\ref{235}), (\ref{236}) correspond to model of the
rotor and can be equally obtained from the action
\begin{eqnarray}\label{238}
S=\int d\tau\left[\frac{1}{2}\dot x^2 +\phi(x^2-1)\right].
\end{eqnarray}
Thus we have demonstrated canonical equivalence of the models
(\ref{232}), (\ref{233}) and (\ref{235}), (\ref{236}).
They correspond to different choices of physical variables in the
underlying action (\ref{230}). Equivalently, they are related by the
change of variables $v_i=\epsilon_{ijk}x_jp_k$.

Resolving the remaining constraints from Eq.(\ref{236}), it is not
difficult to find the canonical variables of the model
$x_a=\tilde x_a, ~  p_a=\tilde p_a-\frac{(\tilde x \tilde p)}
{1+\tilde x^2}\tilde x_a, ~ a=1, 2$,
which obey $\{\tilde x_a, \tilde x_b\}=\{\tilde p_a, \tilde p_b\}=0,
~ \{\tilde x_a, \tilde p_b\}=\delta_{ab}$.
Dynamics is governed by the nonlinear equations
$\dot{\tilde x}_a=\tilde p_a+ \frac{(\tilde x \tilde p)}{1-\tilde x^2}
\tilde x_a, ~
\dot{\tilde p}_a=0$.
Let us point out that the relation established between the particle with
a monopole and the
rotor allows one to construct NC quantum mechanics corresponding to
the geometry given in Eq.(\ref{232}),
with the nontrivial potential (\ref{234}), following the same procedure
as in Sect. 2. Ssince the bracket kernel of (\ref{235})
is degenerated,
see Eq.(\ref{236}), the star product constructed using all six variables
turns out to be nonassociative [25]. This subject will be discussed
elsewhere.

{\bf $(v_i, ~ p_i)$-representation.} For these variables the algebra is
$(J_i\equiv\epsilon_{ijk}v_jp_k)$
\begin{eqnarray}\label{241}
\{v_i, v_j\}=\epsilon_{ijk}v_k, \quad
\{v_i, p_j\}=-\frac{1}{p^2}(v_iJ_j-v_jJ_i), \quad
\{p_i, p_j\}=-\epsilon_{ijk}v_k,
\end{eqnarray}
while the equations of motion are similar to $(x, v)$-representation
\begin{eqnarray}\label{242}
\dot v_i=0, \quad \dot p_i=2\epsilon_{ijk}v_jp_k, \quad
v_i^2=p_i^2, \quad p_iv_i=0.
\end{eqnarray}
Comparing this representation with $(x, ~ p)$-representation one
observes the same property as for NC sphere: transition from commutative
description (\ref{235}) to NC description (\ref{241}) implies trivial
dynamics for space variables in the latter representation.

Thus we have presented the Lagrangian formulations for a
particle on the noncommutative sphere (\ref{201}) as well as for a
particle on the commutative sphere with a monopole at the
center (\ref{230}), the latter is shown to be canonically equivalent
to the model of rotor.
In both cases the desired algebraic structure (\ref{222}), (\ref{232})
arises as the Dirac bracket corresponding to
the second class constraints
presented in the model. After introduction of the Dirac bracket, the
constraints can be used to represent part of variables through the
remaining ones. There exist several ($SO(3)$ covariant)
possibilities to choose the basic variables, which leads to different
representations for the two models. In both cases there is the
``commutative representation'' which is appropriate for determining
the canonical variables starting from the known constraint system.
Using relation
between NC and commutative representations one is able to construct
quantum mechanics which corresponds to the NC representation.

In conclusion, let us comment on possible generalizations of the
model (\ref{201}). One possibility is to consider immersion of the model
into a locally invariant system. Let us omit the term $\phi(x^2-1)$
in the action (\ref{1}). Then the formulation involves one first class
constraint which corresponds to the local symmetry
$\delta x_i=\gamma v_i$. Thus, one is able now to consider different
gauges of the model ($x^2-1=0$ and $v^2-1=0$ are equally
admissible now). We suggest that it can give unified description of
the three models considered in this work. Other possibility may be
NC quantum mechanics on three-dimensional plane. To construct it, one
needs to modify the action (\ref{201}) in such a way that only primary
constraints of the type (\ref{202}) are generated and form the second
class system. These problems will be considered elsewhere.

\section{Acknowledgments.}
Author thanks INFN for the financial support and hospitality during
his stay in Frascati, Italy, where part of this work was done.

\end{document}